# The Intrinsic Shapes of Stellar Systems

Barbara S. Ryden [1]

Department of Astronomy, The Ohio State University,

174 W. 18th Ave., Columbus, OH 43210

## ABSTRACT

I compute the estimated distribution function $f(q)$ for the apparent axis ratio $q$ of various types of stellar systems, using a nonparametric kernel method. I then invert $f(q)$ to find the distribution of intrinsic axis ratios, using two different hypotheses: first, that the stellar systems are all oblate, and second, that they are all prolate. The shapes of globular clusters in our own galaxy are consistent, at the 99% confidence level, with both the oblate and prolate hypothesis. The shapes of dwarf galaxies in the Virgo cluster are consistent, at the 99% confidence level, with the prolate hypothesis, but inconsistent with the oblate hypothesis. The shapes of star clusters in the Large Magellanic Cloud, of ordinary elliptical galaxies, of brightest cluster ellipticals, and of galaxy clusters are all inconsistent, at the 99% confidence level, with both the oblate and prolate hypotheses. The globular clusters in our galaxy are older than their half-mass relaxation time, and are most likely rotationally flattened oblate spheroids. The other stellar systems considered are generally younger than their half-mass relaxation time, and thus are triaxial bodies flattened by anisotropy of their velocity dispersion.

*Subject headings:* galaxies: clustering – galaxies: elliptical and lenticular, cD – galaxies: structure – globular clusters: general

## 1. Introduction

Star clusters, galaxies, and galaxy clusters can only be seen in projection against the sky. Astronomers can only attempt to deduce the three-dimensional properties of these stellar systems from their two-dimensional projected properties. Some information is lost in the process of projection, however. For instance, using only the two-dimensional surface photometry, it is impossible to state the intrinsic three-dimensional shape of a given stellar system.

---

Elliptical galaxies, for example, have isophotes which are well approximated as ellipses. The shape of an ellipse is specified by its axis ratio $q$, where $0 \leq q \leq 1$. Since the isophotes of an elliptical galaxy are nearly elliptical, its isoluminosity surfaces are generally modeled as ellipsoids. A stellar system whose isoluminosity surfaces are similar, concentric ellipsoids without axis twisting, will have projected isophotes which are similar, concentric ellipses without axis twisting (Contopoulos 1956, Stark 1977). The apparent axis ratio $q$ of the projected ellipse depends on the viewing angle and on the intrinsic axis ratios $\beta$ and $\gamma$ of the ellipsoid. I choose $\beta$ to be the ratio of the intermediate axis to the long axis, and $\gamma$ to be the ratio of the short axis to the long axis; thus, $0 \leq \gamma \leq \beta \leq 1$.

Hubble (1926), in his classification scheme for galaxies, was forced to classify elliptical galaxies according to their apparent axis ratios, since their intrinsic axis ratios are unknown. In general, even a perfect knowledge of the distribution of apparent axis ratios, $f(q)$, for all elliptical galaxies doesn't uniquely determine the distribution of intrinsic axis ratios, $f(\beta, \gamma)$. However, by making the assumption that elliptical galaxies are oblate spheroids, with $\beta = 1$, and that they are randomly oriented with respect to our own galaxy, Hubble was able to invert the observed distribution of apparent axis ratios to find the distribution of intrinsic axis ratios. For half a century after the work of Hubble, the standard assumption, in the absence of evidence to the contrary, was that elliptical galaxies were oblate spheroids, flattened by rotation. Sandage, Freeman, & Stokes (1970) in their analysis of the intrinsic flattening of galaxies, assumed that all elliptical galaxies are oblate.

The realization that elliptical galaxies are not rotationally flattened (Bertola & Capaccioli 1975; Illingworth 1977) led astronomers to abandon the assumption that ellipticals are necessarily oblate. The shapes of elliptical galaxies have been reanalyzed with the hypotheses that they are intrinsically prolate or triaxial, rather than oblate (Binney 1978; Benacchio & Galletta 1980; Binggeli 1980; Binney & de Vaucouleurs 1981). Recently, careful surface photometry of ellipticals has revealed a lack of E0 galaxies – that is, galaxies which are nearly circular. The distribution $f(q)$ of apparent shapes peaks at $q \sim 0.8$ (Benacchio & Galletta 1980; Fasano & Vio 1991; Franx, Illingworth, & de Zeeuw 1991; Lambas, Maddox, & Loveday 1992). The scarcity of E0 galaxies cannot be reproduced by a population of randomly oriented spheroids; it can easily be produced, however, in models where the galaxies are triaxial ellipsoids (Lambas et al. 1992; Ryden 1992; Tremblay & Merritt 1995). Kinematic information, where it is available, also contradicts the simple picture of elliptical galaxies as spheroids rotating about their axis of symmetry (Franx et al. 1991; Statler 1994a,b; Statler & Fry 1994).

Once elliptical galaxies were discovered to be triaxial, studies were made of the intrinsic shapes of other stellar systems as well. The apparent shapes of globular clusters associated with our own galaxy are consistent with their all being oblate spheroids (White & Shawl 1987; Han & Ryden 1994). Clusters in the Large Magellanic Cloud, by contrast, show the paucity of nearly circular systems which is a distinguishing characteristic of a population of triaxial bodies (Frenk & Fall 1982; van den Bergh & Morbey 1984; Kontizas et al. 1989; Han & Ryden 1994). The apparent shapes of dwarf elliptical galaxies are also best explained by a parent population of intrinsically triaxial objects (Ryden & Terndrup 1994; Binggeli & Popescu 1995). At the other end of the luminosity function for galaxies, brightest cluster galaxies in Abell clusters are also best explained by a triaxial population (Porter, Schneider, & Hoessel 1991; Ryden, Lauer, & Postman 1993).

Expressing the shape of a cluster of galaxies with a single number $q$ is a far riskier undertaking, since clusters are raggedy objects, undergoing the irregular accretion of clumps of galaxies. Despite the presence of substructure, apparent axis ratios for clusters have been estimated using a variety of techniques. Early samples, containing small numbers of clusters, were judged to be consistent with both the oblate and the prolate hypothesis (Carter & Metcalfe 1980; Binggeli 1982; Di Fazio & Flin 1988). More recent studies, based on samples of hundreds of clusters, are able to reject the hypothesis that clusters of galaxies are oblate, but are consistent with the hypotheses that they are prolate or triaxial (Plionis, Barrow, & Frenk 1991; Salvador-Solé & Solanes 1993; de Theije, Katgert, & van Kampen 1995).

Statements about the intrinsic shapes of stellar systems must be statistical in nature. Astronomers do not exactly know the distribution $f(q)$ of axis ratios for a given type of stellar system. Instead, they have a finite sample drawn from the parent population $f(q)$. If, purely by chance, the sample happens to contain very few circular systems, I might falsely reject the oblate hypothesis if I do not properly take into account the statistics of the problem. In this paper, I reject or accept, at a known confidence level, the null hypothesis that a given population of stellar systems consists of randomly oriented oblate spheroids. I similarly accept or reject the hypothesis that they are randomly oriented prolate spheroids. To accomplish this, I make a nonparametric estimate $\hat{f}(q)$ of the distribution of axis ratios, and mathematically invert $\hat{f}(q)$ to find $\hat{N}_o(\gamma)$ and $\hat{N}_p(\gamma)$, the estimated distribution of intrinsic axis ratios for a population of (respectively) oblate and prolate spheroids. Confidence intervals are placed on the estimators $\hat{f}$, $\hat{N}_o$, and $\hat{N}_p$ by repeatedly performing bootstrap resampling of the original data, and creating new estimates from each sample. The spread in the bootstrap estimates of $f$ at a given value of $q$ and of $N_o$ and $N_p$ at a given value of $\gamma$ provide confidence intervals for the nonparametric estimates of these functions.

In section 2, I give a brief review of the nonparametric kernel estimators which I use in this paper. In sections 3, 4, and 5, I find the kernel estimates of the intrinsic axis ratios of star clusters, of galaxies, and of clusters of galaxies. Although globular star clusters in our own galaxy can be oblate or prolate, for the other stellar systems which I examine, I can reject both the oblate and the prolate hypotheses at a confidence level of 90% or higher. The frequency of triaxiality, and its implications for the formation and evolution of stellar systems, is discussed in section 6.

## 2. Methods

Given a sample $q_1, q_2, \ldots, q_N$ of measured axis ratios, the kernel estimator of the distribution $f(q)$ is

$$\hat{f}(q) = \frac{1}{Nh} \sum_{i=1}^{N} K\left(\frac{q - q_i}{h}\right) , \qquad (1)$$

where $K$ is the kernel function. For my purposes, I wanted a smooth, differentiable kernel, so I chose a Gaussian:

$$K(x) = \frac{1}{\sqrt{2\pi}} e^{-x^2/2} . \qquad (2)$$

General reviews of kernel estimators are given by Silverman (1986) and Scott (1992). Some applications to astronomical problems are described by Vio et al. (1994) and by Merritt & Tremblay (1994).

Choosing the kernel width $h$ requires some care. A large value of $h$ will result in an estimated function $\hat{f}$ which is smooth but biased. A small value of $h$ will result in a function which is noisy but unbiased; in an average sense, it will follow the true function $f$. One objective method of selecting the kernel width is to use the value of $h$ which minimizes the integrated mean square error, defined as the expectation value of the integral

$$\int [\hat{f}(q) - f(q)]^2 dq . \qquad (3)$$

For distributions which are reasonably smooth, and not strongly skewed, the value of $h$ which minimizes the integrated mean square error is well approximated by the relation (Silverman 1986; Vio et al. 1994)

$$h = 0.9 A N^{-0.2} , \qquad (4)$$

where $A$ is the smaller of the standard deviation of the sample and the interquartile range divided by 1.34. In this paper, I use the above approximation for the value of $h$. This kernel width yields estimates of $f$ which are sufficiently smooth to satisfy my (highly subjective) esthetic criteria, but which are not biased with respect to noisier estimates which I computed using $h$ half as large.

Because $q$ is limited, by definition, to the range $0 \leq q \leq 1$, attention must be paid to boundary conditions. A estimate of the function $f(q)$ which predicts $f \neq 0$ for $q < 0$ or $q > 1$ is not physically acceptable. I apply reflective boundary conditions at the boundaries $q = 0$ and $q = 1$ (see Silverman [1986] for a discussion of reflective boundary conditions). This is done by replacing the Gaussian kernel $K$ of equation (2) with the kernel

$$K_{ref}(q, q_i, h) = K\left(\frac{q - q_i}{h}\right) + K\left(\frac{q + q_i}{h}\right) + K\left(\frac{2 - q - q_i}{h}\right) . \qquad (5)$$

In essence, the Gaussian tails which extend to the left of $q = 0$ and to the right of $q = 1$ are being folded back into the interval $0 \leq q \leq 1$. The use of reflective boundary conditions ensures the proper normalization

$$\int_0^1 \hat{f}(q) dq = 1 \qquad (6)$$

as long as $h \ll 1$. One drawback of using reflective boundary conditions is that it compels

$$\left. \frac{d\hat{f}}{dq} \right|_{q=0} = 0, \quad \left. \frac{d\hat{f}}{dq} \right|_{q=1} = 0 , \tag{7}$$

a condition which is not necessarily true for the authentic function $f(q)$. The reader is therefore cautioned not to believe the shape of $\hat{f}$ within a distance $\sim h$ of the boundaries at $q = 0$ and $q = 1$. Reflective boundary conditions are not the only possible choice. For instance, Vio et al. (1994) deal with the boundaries by ignoring them, and permitting $q$ to take on values $q < 0$ and $q > 1$. Tremblay & Merritt (1995), as an alternative to reflective boundary conditions, also use the transformation $q' = -\ln(1-q)$, which moves the boundary at $q = 1$ to infinity. In practice, it is the boundary at $q = 1$ which is the troublesome one, since $f(q)$ for real stellar systems dwindles to zero before reaching $q = 0$. In this paper, I restrict myself to the robust reflective boundary conditions.

Suppose that the stellar systems which we are examining are all oblate spheroids, randomly oriented with respect to us. From $\hat{f}(q)$, the estimated distribution $\hat{N}_o(\gamma)$ of intrinsic axis ratios can be computed from the relation

$$\hat{N}_o(\gamma) = \frac{2\gamma\sqrt{1-\gamma^2}}{\pi} \int_0^\gamma \frac{d}{dq}\left[\hat{f}/q\right] \frac{dq}{\sqrt{\gamma^2 - q^2}} . \tag{8}$$

The above equation assumes that $\hat{f}(0) = 0$. If, by contrast, the stellar systems are assumed to be prolate, the estimated distribution $\hat{N}_p(\gamma)$ of intrinsic axis ratios is given by the relation

$$\hat{N}_p(\gamma) = \frac{2\sqrt{1-\gamma^2}}{\pi\gamma} \int_0^\gamma \frac{d}{dq}\left[q^2\hat{f}\right] \frac{dq}{\sqrt{\gamma^2 - q^2}} . \tag{9}$$

To be physically meaningful, $\hat{N}_o$ and $\hat{N}_p$ must be non-negative over their entire range $0 \leq q \leq 1$. Several previous investigators have used the iterative method of Lucy (1974) to find $\hat{N}_o$ and $\hat{N}_p$ for elliptical galaxies (Noerdlinger 1979; Binney & de Vaucouleurs 1981; Fasano & Vio 1991; Franx et al. 1991; Vio et al. 1994) and for galaxy clusters (de Theije et al. 1995). Lucy's method ensures that the solution found is smooth and positive – even if a smooth, positive distribution for $\gamma$ gives a poor fit to the observed distribution of $q$. My approach to finding $\hat{N}_o$ and $\hat{N}_p$ is somewhat different. I start with the null hypothesis that all objects are oblate. After computing $\hat{f}(q)$, I find the corresponding $\hat{N}_o(\gamma)$ by doing a numerical integration of equation (8), letting $\hat{N}_o$ fall below zero if that is the correct solution from a mathematical viewpoint. In this way, I am attempting to disprove the null hypothesis by a reductio ad absurdum argument: I assume that all objects are oblate, and show that my assumption results in the absurd conclusion that $\hat{N}_o$ falls below zero. In order to test the null hypothesis that all objects are prolate, I use a similar procedure, using equation (9) to find $\hat{N}_p$.

In order to exclude the oblate or prolate hypothesis at some statistical confidence level, I must show that the excursion of $\hat{N}_o$ or $\hat{N}_p$ below zero is statistically significant. The estimated function $\hat{f}$ contains errors both because of the finite sample size and because of errors in measuring $q$ for the individual stellar systems. The error due to finite sample size can be estimated by bootstrap resampling of the data set. From the original data set $q_1$, $q_2$, ..., $q_N$, for a given sample of $N$ stellar systems, I draw, with replacement, a new set of $N$ data points. I then use these points to create a new estimate $\hat{f}$, using equation (1). From this 'bootstrap estimate' (to use the terminology of Merritt & Tremblay 1994), I compute the estimated values $\hat{N}_o$ and $\hat{N}_p$. After creating a large number of these bootstrap estimates for $N_o$ and $N_p$, confidence intervals can be placed on the original estimates. For instance, at each value of $\gamma$, 98% confidence intervals can be placed on $\hat{N}_o$ (or $\hat{N}_p$) by finding the values of $\hat{N}_o$ (or $\hat{N}_p$) such that 1% of the bootstrap estimates lie above the upper confidence limit and 1% lie below the lower confidence limit. If this upper confidence limit drops below zero for any value of $\gamma$, the hypothesis that all objects are randomly oriented and oblate (or prolate) can be rejected at the 99% one-sided confidence level. For each data set examined in this paper, I made 800 bootstrap resamplings, in order to have sufficiently accurate measurements of the confidence levels at high confidence levels.

The confidence bands derived from bootstrap resampling represent the variance due to the finite sample size (Scott 1992, Merritt & Tremblay 1994). An additional source of error in $\hat{N}_o$ and $\hat{N}_p$ is the error which is inevitably present in the measured values of the apparent axis ratio $q$. Suppose that the measurement of $q_i$ for a given stellar system has an estimated error $\sigma_i$. If all the values of $\sigma_i$ for a data set are smaller than the kernel width $h$, then the effect of measurement errors can be ignored. For the data sets examined in this paper, however, the errors $\sigma_i$ area comparable in size to $h$, and thus may have a noticeable effect on the estimated values of $\hat{N}_o$ and $\hat{N}_p$. For instance, White & Shawl (1987) and Kontizas et al. (1989) both find typical formal errors of $\sigma \sim 0.03$ in measuring $q$ for globular clusters; this is slightly larger than the kernel width $h = 0.022$ which I apply to their data sets. The other measurements of $q$ which I took from the literature are not accompanied by published errors. The effect of Gaussian errors with a uniform standard deviation $\sigma$ can be modeled by replacing the kernel width $h$ given by equation (4) with a width

$$h' = \sqrt{h^2 + \sigma^2} \ . \tag{10}$$

For each set of stellar systems, I compared the values of $\hat{N}_o$ and $\hat{N}_p$ found first with $\sigma = 0$ and then with $\sigma = 0.03$. In most cases, the addition of a $\sigma = 0.03$ Gaussian error does not affect whether the oblate or prolate hypothesis can be rejected at a high confidence level. For the sample of elliptical galaxies described in section 4, however, the prolate hypothesis can be rejected at the 90% confidence level when $\sigma = 0$, but cannot be rejected at the 90% confidence level when $\sigma = 0.03$.

The presence of modestly sized errors ($\sigma \lesssim h$) with a Gaussian distribution does not greatly affect the values of $\hat{N}_o$ and $\hat{N}_p$. A complicating factor is that the errors in $q$ do not always have a Gaussian distribution. A stellar system which has a true value of $q = 1$ will have a measured value of $q \leq 1$. Galaxies or clusters which are precisely circular, in other words, will be measured to be slightly flattened. This systematic decrease in $q$ when $q \sim 1$ can have a significant effect on the deduced distribution of intrinsic shapes, particularly when a large fraction of the observed systems have $q \gtrsim 1 - \sigma$, where $\sigma$ is a typical error for $q$ in nearly circular systems. For instance, Franx & de Zeeuw (1992) deduced the intrinsic ellipticity $\epsilon$ of disk galaxies, starting with the measured apparent axis ratios of $\sim 600$ disks. Assuming Gaussian errors, they found the best fitting model had $\epsilon = 0.06$; with non-Gaussian errors, the best fitting model had $\epsilon = 0$. However, the disks studied by Franx & de Zeeuw were drawn from a sample (Grosbøl 1985) selected to have $q > 0.56$, and which has a roughly uniform distribution in $q$. By contrast, the stellar systems studied in this paper, with the exception of the globular clusters in Figure 1, show a marked lack of systems with measured values of $q \gtrsim 0.9$ – see Figures 2 through 6. This lack of nearly circular systems occurs over too broad a range of $q$ to be entirely attributed to non-Gaussian errors in the range $q \gtrsim 1 - \sigma$. For the remainder of this paper, I will simply assume that any errors in $q$ are sufficiently small, in comparison to $h$, to be ignored.

## 3. Globular Clusters

The globular clusters associated with our own galaxy are nearly circular in projection – hence the name 'globular'. The small deviations from sphericity, however, can be measured. White & Shawl (1987), in a sample of 99 globular clusters, found a mean axis ratio $<q> = 0.93$; only five of the clusters had $q < 0.8$. White and Shawl measured the axis ratio of their clusters at an average radius of $0.7 r_h$, where $r_h$ is the half-mass radius of the cluster. Within $r_h$, the distortion due to the tidal field of our galaxy is negligible; the observed flattening must then be due to rotation or velocity anisotropy of the globular clusters.

The estimated distribution $\hat{f}(q)$ of axis ratios for the sample of White & Shawl is shown as the solid line in the upper panel of Figure 1. A kernel width of $h = 0.022$ is used. The dashed lines in every Figure indicate the 80% confidence band, and the dotted lines indicate the 98% confidence band, as found by bootstrap resampling of the original data set. The data are seen to be consistent with a monotonically increasing distribution for $q$, peaking at $q = 1$. The estimated distribution $\hat{N}_o(\gamma)$ of axis ratios, given the oblate hypothesis, is shown in the middle panel of Figure 1. The best-fitting kernel estimate (the solid line) is positive everywhere, so the observed distribution of intrinsic shapes is consistent with the hypothesis that the globular clusters are randomly oriented oblate spheroids. Given the width of the confidence bands, there is no evidence that the distribution of $\gamma$ is multimodal. The distribution $\hat{N}_p(\gamma)$, shown in the bottom panel of Figure 1, is also positive everywhere. From their apparent shapes alone, it's impossible to state whether the globular clusters in our galaxy are oblate or prolate. If the clusters are all oblate, their estimated mean axis ratio is

$$<\gamma>_O \equiv \int_0^1 \hat{N}_o(\gamma) d\gamma = 0.881 \ . \tag{11}$$

If they are all prolate, their estimated mean axis ratio is

$$<\gamma>_P \equiv \int_0^1 \hat{N}_p(\gamma)d\gamma = 0.889 \ . \qquad (12)$$

Kontizas et al. (1989) measured the apparent axis ratios at $r_h$ of 49 bright clusters in the Large Magellanic Cloud (LMC). These star clusters are comparable in luminosity to the globular clusters in our own galaxy, but they are, on average, considerably more flattened. Their average axis ratio is $<q> = 0.84$, and the roundest cluster in the sample of Kontizas et al. has $q = 0.94$. The estimated distribution of $q$, as shown in the upper panel of Figure 2, is consistent with a unimodal distribution peaking at $q = 0.86$, and with a distinct lack of nearly circular clusters. When the LMC clusters are assumed to be oblate, the paucity of circular clusters results, as seen in the middle panel of Figure 2, in a best estimate for $N_o$ becomes negative as $\gamma \to 1$. The 98% confidence band falls below zero for $\gamma > 0.906$. The best estimate for $N_p$, under the prolate hypothesis, also becomes negative, as seen in the bottom panel of Figure 2. Here, the 98% confidence band falls below zero for $\gamma > 0.911$. Thus, both the oblate and the prolate hypothesis can be firmly rejected at the 99% one-sided confidence level.

The shapes of globular clusters in the Milky Way are consistent with their being oblate or prolate spheroids. The shapes of the analogous clusters in the LMC, by contrast, can only be explained by their being triaxial ellipsoids (or, implausibly, by a conspiracy which orients them edge-on with respect to an observer on Earth). The difference in shape between the two populations may be explained by the difference in their ages. Globular clusters in our galaxy have ages $t \sim 1.5 \times 10^{10}$ yr. Clusters in the LMC have a much wider range of ages, and are younger on average, with $10^6$ yr $\lesssim t \lesssim 10^{10}$ yr (Elson, Fall, & Freeman 1987). Stellar systems which form by anisotropic collapse will generally have anisotropic velocity dispersions once violent relaxation brings the system into dynamic equilibrium (Aarseth & Binney 1978). During this epoch, after violent relaxation but before two-body relaxation, the stellar system will be a triaxial ellipsoid flattened by anisotropic dispersion. After a time roughly equal to $t_{rh}$, the two-body relaxation time computed within the half-mass radius of the system, the stellar system will relax to a state in which its velocity dispersion is nearly isotropic (Fall & Frenk 1983). During this epoch, the system will be an oblate spheroid flattened by rotation, with an axis ratio given by the relation (Fall & Frenk 1983)

$$\left(\frac{v}{\sigma}\right)^2 = \frac{(1+2\gamma^2)\cos^{-1}\gamma - 3\gamma\sqrt{1-\gamma^2}}{\gamma\sqrt{1-\gamma^2} - \gamma^2\cos^{-1}\gamma} \ , \qquad (13)$$

where $v$ is the rotation velocity and $\sigma$ is the one-dimensional velocity dispersion. Finally, evaporation modifies the shape of an isolated stellar system over a time scale $\sim 40 t_{rh}$ (Fall & Frenk 1985; Davoust & Prugniel 1990). Stars moving in the same sense as the system's rotation are more likely to escape; hence, the loss of stars decreases the rotation velocity and the flattening of the cluster.

Globular clusters in our own galaxy mostly have half-mass relaxation times lying in the range $10^8$ yr $\lesssim t_{rh} \lesssim 10^{10}$ yr (Ostriker, Spitzer, & Chevalier 1972). Thus, the average globular cluster in our galaxy should be a rotationally flattened oblate spheroid with an isotropic velocity dispersion. Nearby globular clusters, for which accurate kinematic data, are flattened along their kinematic minor axis, and can be adequately modeled as rotationally flattened spheroids (Davoust 1986). Moreover, there exists a correlation between the apparent axis ratio and the relaxation time scale for globular clusters in our galaxy; clusters with shorter relaxation times tend to be rounder (Davoust & Prugniel 1990). In the LMC, a sample of ten young clusters examined by Elson et al. (1987) all had ages $t \leq 3 \times 10^8$ yr but half-mass relaxation times $t_{rh} \geq 10^9$ yr. All but the oldest clusters in the LMC should still have anisotropic velocity dispersions and triaxial shapes, a conclusion which agrees with the observed distribution of apparent shapes.

## 4. Galaxies

The luminosity function of dwarf elliptical (dE) galaxies overlaps that of globular clusters. In the luminosity range $-10 \lesssim M_B \lesssim -8$, one can find both dim dwarf ellipticals and bright globular clusters. However, dwarf ellipticals and globular clusters are distinguished by having extremely different surface brightness. A typical central surface brightness for a bright globular cluster is $\mu_{0V} \sim 16\,\text{mag}\,\text{arcsec}^{-2}$; for a dim dwarf elliptical, it can be as low as $\mu_{0V} \sim 26\,\text{mag}\,\text{arcsec}^{-2}$ (Kormendy 1985). The low stellar density within dE's implies that their two-body relaxation time is much longer than the age of the universe. For a stellar system with $N$ stars, each of mass $\sim 1 M_\odot$, the half-mass relaxation time is (Goodman & Lee 1989)

$$t_{rh} \approx 5 \times 10^9 \,\text{yr} \left(\frac{M}{10^6 M_\odot}\right)^{1/2} \left(\frac{r_h}{10\,\text{pc}}\right)^{3/2} \left(\frac{13}{\ln[0.4N]}\right). \quad (14)$$

A dwarf elliptical with $r_h \sim 1$ kpc will have a relaxation time 1000 times longer than that of a globular cluster with the same number of stars but with $r_h \sim 10$ pc. Dwarf elliptical galaxies, with half-mass relaxation times in the range $t_{rh} \gtrsim 10^{11}$ yr, will not have had time to undergo two-body relaxation; if they were born triaxial, they will still be triaxial today.

Binggeli & Cameron (1993; hereafter BC) computed the axis ratio $q$ for dwarf galaxies in the Virgo cluster by digitizing photographic plates and fitting elliptical isophotes at the surface brightness level $\mu_B = 26\,\text{mag}\,\text{arcsec}^{-2}$. For a subsample of 170 dwarf galaxies classified by Binggeli & Cameron as 'dE', 'dE:', or 'd:E', the best-fitting estimate $\hat{f}$ for the distribution of $q$ is shown in the upper panel of Figure 3. Dwarf ellipticals are more flattened, on average, than globular clusters; the BC dwarf ellipticals have an average axis ratio $<q>= 0.73$. Inversion of $\hat{f}$ to find $\hat{N}_o$, as displayed in the middle panel of Figure 3, shows that the oblate hypothesis can be rejected at the 99% one-sided confidence level for this sample of dwarf ellipticals. Again, the lack of apparently circular galaxies cannot be explained by a population of randomly oriented oblate spheroids. The *prolate* hypothesis, though, cannot be rejected at the 99% confidence level for this sample; see the bottom panel of Figure 3. If the dwarf ellipticals in the BC sample are prolate, then $\hat{N}_p$ yields an average intrinsic axis ratio $<\gamma>_P= 0.65$.

Ryden & Terndrup (1994; hereafter RT) performed R-band CCD imaging of 70 dE's in the Virgo cluster, including 26 galaxies also in the BC sample. For each galaxy, RT computed a luminosity-weighted mean axis ratio $\overline{q}$. The galaxies in the RT sample are slightly flatter, on average, than the galaxies in the BC sample, having $<\overline{q}> = 0.70$. As measured by a Kolmogorov-Smirnov (KS) test, however, the distribution of $\overline{q}$ found by RT is not significantly different from the distribution of $q$ found by BC; the KS probability from comparing the two samples is $P_{KS} = 0.11$. When I compute $\hat{f}$ for the RT sample and invert it to find $\hat{N}_o$ and $\hat{N}_p$, I find that the oblate hypothesis can be rejected at the 90% confidence level (but not at the 99% confidence level) and that the RT data are fully consistent with the prolate hypothesis. If the dwarf ellipticals in the RT sample are prolate, then $\hat{N}_p$ yields $<\gamma>_P = 0.61$.

Binggeli & Cameron (1993) also measured the apparent axis ratio $q$ for 15 dS0 galaxies in the Virgo cluster. The distribution of $q$ for these galaxies is consistent with the oblate hypothesis; the best fitting $\hat{N}_o$ under the oblate hypothesis peaks at $\gamma \sim 0.33$, with an average axis ratio $<\gamma>_O = 0.40$. However, the distribution of $q$ for these galaxies is also consistent with the prolate hypothesis, with $<\gamma>_P = 0.56$. In the absence of kinematic information, it is impossible to state whether the dS0 galaxies are oblate or prolate or triaxial. Moreover, a sample of 15 dS0 galaxies is too small to make definitive statistical statements. A KS comparison of the 170 dE galaxies in the BC sample and the 15 dS0 galaxies yields a probability $P_{KS} = 0.14$. At confidence levels greater than 86%, I cannot reject the hypothesis that the dE sample and the dS0 sample are drawn from the same distribution of shapes.

The apparent shapes of giant elliptical (E) galaxies differ from those of dwarf elliptical (dE) galaxies; on average, dE's are flatter than E's (Ryden & Terndrup 1994). However, giant ellipticals, like dwarf ellipticals, are characterized by a lack of nearly circular E0 galaxies. Outside their central cores, the two-body relaxation time for elliptical galaxies is longer than the age of the universe. Thus, elliptical galaxies should retain the anisotropic velocity distribution which was created during their formation, whether the formation process was nonspherical collapse (Aarseth & Binney 1978) or a merger of progenitor galaxies (Barnes & Hernquist 1992).

The top panel of Figure 4 illustrates the estimated distribution of apparent axis ratios for 165 bright elliptical galaxies observed by Djorgovski (1985). The luminosity-weighted axis ratios $\overline{q}$ for these galaxies were computed by Ryden (1992). Similar distributions for $q$, peaking at $q \sim 0.8$, and declining as $q \to 1$, have been found by Benacchio & Galletta (1980), Fasano & Vio (1991), Franx et al. (1991), and Lambas et al. (1992). Inverting $\hat{f}$ to find $\hat{N}_0$, as shown in the central panel of Figure 4, reveals that the apparent shapes of E galaxies are inconsistent with the oblate hypothesis. The 98% confidence band for $\hat{N}_o$ lies below zero when $\gamma > 0.949$. According to the criteria I have been using, the prolate hypothesis as well as the oblate hypothesis can be rejected at the 99% one-sided confidence level. The 98% confidence band for $\hat{N}_p$, as shown in the bottom panel of Figure 4, falls below zero when $\gamma > 0.966$. However, I used a kernel width $h = 0.033$ when constructing $\hat{f}$, and one should not take seriously the shape of $\hat{f}$, $\hat{N}_o$, and $\hat{N}_p$ within a distance $\sim h$ of the boundary. To reliably rule out the prolate hypothesis would require a larger sample of elliptical galaxies. Increasing $N$ will decrease the width of the confidence bands, and also slightly decrease the optimum kernel width $h$.

In rich clusters of galaxies, the brightest cluster galaxy (or BCG) is usually a cD galaxy or an unusually bright E galaxy. Because they are of extremely high mass, and are located in the centers of clusters, BCG's should have evolutionary histories which include mergers, cannibalism, and accretion of tidal debris. An ordinary E galaxy (that is, one that's not a BCG) will have a different evolutionary history, with fewer mergers and, in general, fewer interactions of any sort with neighboring galaxies. The still-contentious state of BCG evolution is reviewed by Kormendy & Djorgovski (1989). Intriguingly enough, there is no evidence that the distribution of intrinsic shapes for BCG's differs from the distribution for elliptical galaxies as a whole. Ryden et al. (1993) computed the luminosity-weighted axis ratio $\overline{q}$ for a sample of 119 BCG's in Abell clusters with $cz < 15{,}000 \text{ km s}^{-1}$. As measured by a KS test, the distribution of $\overline{q}$ for BCG's was not significantly different ($P_{KS} = 0.14$) from the distribution of $\overline{q}$ for the 165 ordinary ellipticals in the Djorgovski (1985) sample analyzed by Ryden (1992).

The kernel estimate of $\hat{f}(q)$ for the BCG sample is shown in the top panel of Figure 5. Once again, the steep decline in $\hat{f}$ as $q \to 1$ is the characteristic signature of a population of triaxial ellipsoids. The estimate $\hat{N}_o$, for a hypothesized population of oblate spheroids, is shown in the middle panel of Figure 5. The oblate hypothesis can be rejected at the 99% one-sided confidence level. The 98% confidence band lies below zero for $\gamma > 0.921$. The prolate hypothesis can also be rejected at the 99% confidence level, as the lower panel of Figure 5 reveals. The 98% confidence band for $\hat{N}_p$ falls below zero when $\gamma > 0.935$. A population of purely spheroidal objects, randomly oriented, cannot explained the observed apparent shapes of brightest cluster galaxies. Whatever the evolutionary history of BCG's may be, it has resulted in a family of triaxial galaxies.

## 5. Clusters of Galaxies

A cluster of galaxies is a dynamically young structure. Except in the central region, the crossing time is usually comparable to the cluster's age. Many clusters show substructure: in the number counts of galaxies as projected on the sky (Geller & Beers 1982) and in the distribution of galaxy velocities (Dressler & Schectman 1988; Beers et al. 1991; Bird 1994). A prudent observer of clusters, realizing that a typical cluster of galaxies is not a smooth, virialized, isolated system, would hesitate to characterize the projected shape of a cluster with a single number $q$. However, theorists rush in where observers fear to tread; I have taken computed values of $q$ from the literature, and am using them at face value. The axis ratio of a galaxy can be found by fitting isophotes to the galaxy's surface brightness. The axis ratio of a cluster, by contrast, is usually computed from the position of a few hundred, or at most a few thousand, galaxies within the cluster. The value of $q$ for a given cluster depends on the algorithm used, and how it weights the individual galaxies in the cluster.

One method of computing the axis ratio of a cluster of galaxies is the "tensor" or "principal axis" method (Rhee et al. 1989). The tensor method begins with computing the inertia tensor

$$I_{xx} = \sum x^2/r^2, \quad I_{yy} = \sum y^2/r^2, \quad I_{xy} = \sum xy/r^2 , \tag{15}$$

where the sum is again over all galaxies with measured positions, and $r^2 = x^2 + y^2$. The axis ratio found by the tensor method is $q = \Lambda_-/\Lambda_+$, where

$$\Lambda_\pm = I_{11} + I_{22} \pm \sqrt{(I_{11} + I_{22})^2 - 4(I_{22}^2 - I_{12}^2)} . \tag{16}$$

For observed clusters, $q$ found by the tensor method is usually smaller than $q$ found by the alternative "moment" method (Rhee et al. 1989, de Theije et al. 1995).

Investigators have applied a variety of methods to assorted samples of rich clusters of galaxies. Rhee et al. (1989), using the tensor method to find $q$ for a sample of 107 galaxies, found an average axis ratio of $<q> = 0.81$. Other studies, however, have found that clusters are flatter, on average. Carter & Metcalfe (1980) used a tensor-like method to find $q$ for a sample of 21 clusters; their average axis ratio was $<q> = 0.56$. Binggeli (1982) used a simplified tensor method to compute $q$ for 44 clusters, based on the 50 brightest galaxies in each cluster; he found $<q> = 0.70$ for his sample. Plionis et al. (1991), applying the tensor method to galaxy counts in cells, found $<q> = 0.61$ for a sample of 200 clusters.

Struble & Ftaclas (1994) made a compilation of axis ratio data for a sample of 344 clusters. They drew their data from a variety of published sources. For each source taken individually, they found that the distribution of $q$ was consistent with a Gaussian in $\log q$. For all sources added together, they found $<\log q> = -0.1831$. After correcting the distribution $f(q)$ for each of the published sources to give the same mean value of $\log q$, they found a corrected mean value of $\log q$ for each cluster. Struble & Ftaclas (1994) thus provide us with a value of $q$ for each of 344 clusters. The nonparametric estimate $\hat{f}(q)$ derived from the compilation of Struble & Ftaclas is shown in the top panel of Figure 6. The distribution $\hat{N}_o(\gamma)$ of intrinsic shapes, assuming that the clusters are oblate, is shown in the middle panel. The oblate hypothesis can be rejected at the 99% one-sided confidence level; the 98% confidence band drops below zero for $\gamma > 0.822$. The prolate hypothesis can also be rejected at the 99% one-sided confidence level. As shown in the bottom panel of Figure 6, the 98% confidence band for $\hat{N}_p$ falls below zero in the range $0.893 < \gamma < 0.952$, an interval whose width is equal to $1.8h$.

Applying the same analysis to the Carter & Metcalfe (1980) sample, I am unable, because of the small sample size, to reject either the oblate or the prolate hypothesis at the 99% confidence level. For the Binggeli (1982) sample, I can reject the oblate hypothesis at the 99% confidence level, but not the prolate hypothesis; the same conclusion applies to the Rhee et al. (1989) sample. The sample of Plionis et al. (1991) is statistically indistinguishable from the Struble & Ftaclas (1994) compilation, and can be used to reject both the oblate and prolate hypotheses at the 99% one-sided confidence level.

## 6. Discussion

When I began writing this paper, I wanted to end with the grandiose conclusion, "Almost everything is triaxial." Although such a summary would be overgeneralized, the results presented here show that stellar systems on a wide range of sizes, from star clusters in the LMC to rich Abell clusters of galaxies, have apparent shapes which rule out the hypothesis that they are randomly oriented axisymmetric systems. Of the stellar systems which I examined, only the globular clusters associated with our own galaxy are consistent, at the 99% confidence level, with the hypothesis that they are randomly oriented oblate spheroids. From their apparent shapes alone, the globular clusters may also be prolate spheroids. It is kinematic information which leads us to believe that the globular clusters in our galaxy are rotationally supported oblate spheroids. The globular clusters in our galaxy differ from the other stellar systems in having relaxation times, at their half-mass radius, which are shorter than their ages. Thus, they have had a chance to relax to form rotationally supported oblate spheroids with isotropic velocity dispersions.

The other stellar systems studied – star clusters in the LMC, galaxies, and clusters of galaxies – cannot be reconciled with the oblate hypothesis. In each of the samples examined, there are too few stellar systems whose apparent shapes are nearly circular. The prolate hypothesis is easier to reconcile with the apparent lack of circular systems. A population of randomly oriented hot dogs will contain fewer nearly circular systems than will a population of randomly oriented hamburgers. Thus, for instance, although we can reject the oblate hypothesis at the 99% confidence level for dE galaxies in the Virgo cluster, we cannot reject the prolate hypothesis at the same confidence level. Although a population of prolate spheroids can give a marginally acceptable fit, a population of triaxial ellipsoids will give a much better fit to the data (Ryden & Terndrup 1994; Binggeli & Popescu 1995).

Why are stellar systems (aside from the old, dense globular clusters in our galaxy) triaxial, or possibly prolate in some cases, rather than oblate? The star clusters in the LMC are younger, on average, than the globular clusters in our galaxy; they still retain the anisotropic velocity dispersions and triaxial shapes which they acquired during the formation process. Dwarf elliptical galaxies, elliptical galaxies, and brightest cluster galaxies have stellar densities which are too low (except in the extreme central regions) for two-body relaxation to have taken place. Thus, dE's, E's, and BCG's still retain the triaxial shapes which they had at birth, whether that birth took place by the anisotropic collapse of an initial density perturbation or by the merger of two or more protogalaxies.

Clusters of galaxies are dynamically young, and their current shapes may retain clues about the initial conditions from which large scale structure forms. Their shapes are also a clue to the value of the density parameter, $\Omega_0$. In a universe with a low value of $\Omega_0$, clusters formed early (Richstone, Loeb, & Turner 1992). Clusters in an $\Omega_0 = 0.2$ universe are thus more nearly spherical, on average, than clusters in an $\Omega_0 = 1$ universe (Evrard et al. 1993; de Theije et al. 1995). The shapes of galaxy clusters are also modified by tidal interactions with neighboring

– 14 –

clusters (Salvador-Solé & Solanes 1993).

The ubiquity of triaxiality results from the fact that the triaxial systems are younger than their characteristic two-body relaxation time. They have not yet had time to relax to a rotationally supported axisymmetric state. The shape of triaxial stellar systems is a relic of their formation process and thus can act as useful constraint on theories for the formation of star clusters, of galaxies, and of galaxy clusters.

D. Merritt and B. Tremblay kindly acted as nonparametric mavens. R. Pogge, T. Statler, and C. Han made pertinent comments on the manuscript; T. Lauer made impertinent comments. This work was supported by NSF grant AST-9357396.

– 17 –

Fig. 1.— The top panel shows the nonparametric kernel estimate of the distribution of apparent axis ratios for a sample of 99 globular clusters in our own galaxy (White & Shawl 1987). The middle panel shows the distribution of intrinsic axis ratios, assuming that the clusters are oblate. The bottom panel shows the distribution of intrinsic axis ratios, assuming the clusters are prolate. The solid line in each panel is the best estimate, the dashed lines are the 80% confidence band (found by bootstrap resampling) and the dotted lines are the 98% confidence band. The kernel width is $h = 0.022$.

Fig. 2.— The top panel shows the estimated distribution of apparent axis ratios for a sample of 49 globular clusters in the Large Magellanic Cloud (Kontizas et al. 1989). The middle panel shows the distribution of intrinsic axis ratios, assuming that the clusters are oblate. The bottom panel shows the distribution of intrinsic axis ratios, assuming the clusters are prolate. The solid line in each panel is the best estimate, the dashed lines are the 80% confidence band, and the dotted lines are the 98% confidence band. The kernel width is $h = 0.022$.

Fig. 3.— The top panel shows the estimated distribution of apparent axis ratios for a sample of 170 dwarf elliptical galaxies in the Virgo cluster (Binggeli & Cameron 1993). The middle panel shows the distribution of intrinsic axis ratios, assuming that the dwarfs are oblate. The bottom panel shows the distribution of intrinsic axis ratios, assuming the dwarfs are prolate. The solid line in each panel is the best estimate, the dashed lines are the 80% confidence band, and the dotted lines are the 98% confidence band. The kernel width is $h = 0.047$.

Fig. 4.— The top panel shows the estimated distribution of apparent axis ratios for a sample of 165 elliptical galaxies (Djorgovski 1985; Ryden 1992). The middle panel shows the distribution of intrinsic axis ratios, assuming the ellipticals are oblate. The bottom panel shows the distribution of intrinsic axis ratios, assuming the ellipticals are prolate. The solid line in each panel is the best estimate, the dashed lines are the 80% confidence band, and the dotted lines are the 98% confidence band. The kernel width is $h = 0.033$.

Fig. 5.— The top panel shows the estimated distribution of apparent axis ratios for a sample of 119 brightest cluster galaxies (Ryden et al. 1993). The middle panel shows the distribution of intrinsic axis ratios, assuming the galaxies are oblate. The bottom panel shows the distribution of intrinsic axis ratios, assuming the galaxies are prolate. The solid line in each panel is the best estimate, the dashed lines are the 80% confidence band, and the dotted lines are the 98% confidence band. The kernel width is $h = 0.028$.

Fig. 6.— The top panel shows the estimated distribution of apparent axis ratios for a sample of 344 clusters of galaxies (Struble & Ftaclas 1994). The middle panel shows the distribution of intrinsic axis ratios, assuming the clusters are oblate. The bottom panel shows the distribution of intrinsic axis ratios, assuming the galaxies are prolate. The solid line in each panel is the best estimate, the dashed lines are the 80% confidence band, and the dotted lines are the 98% confidence band. The kernel width is $h = 0.033$.

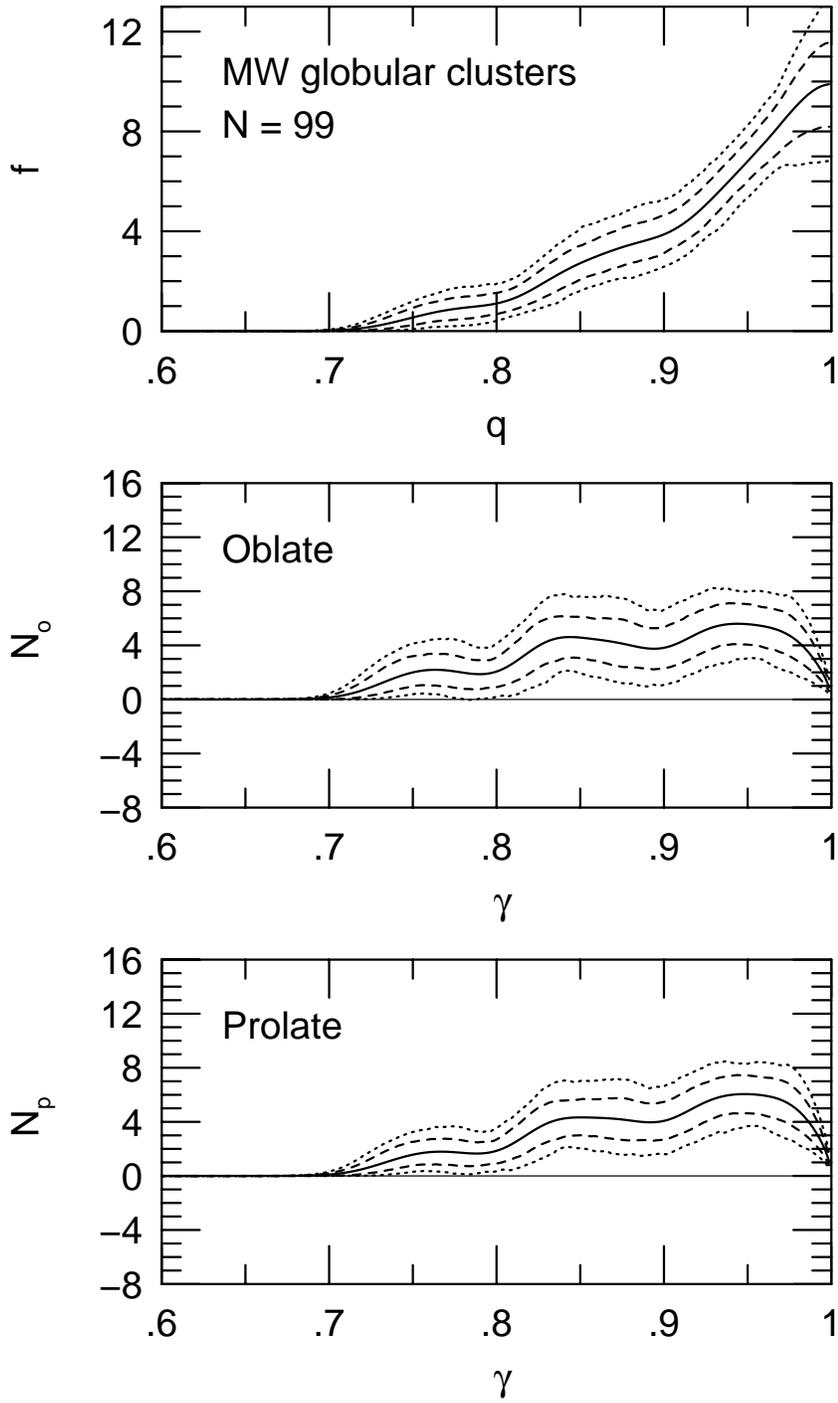

Fig. 1

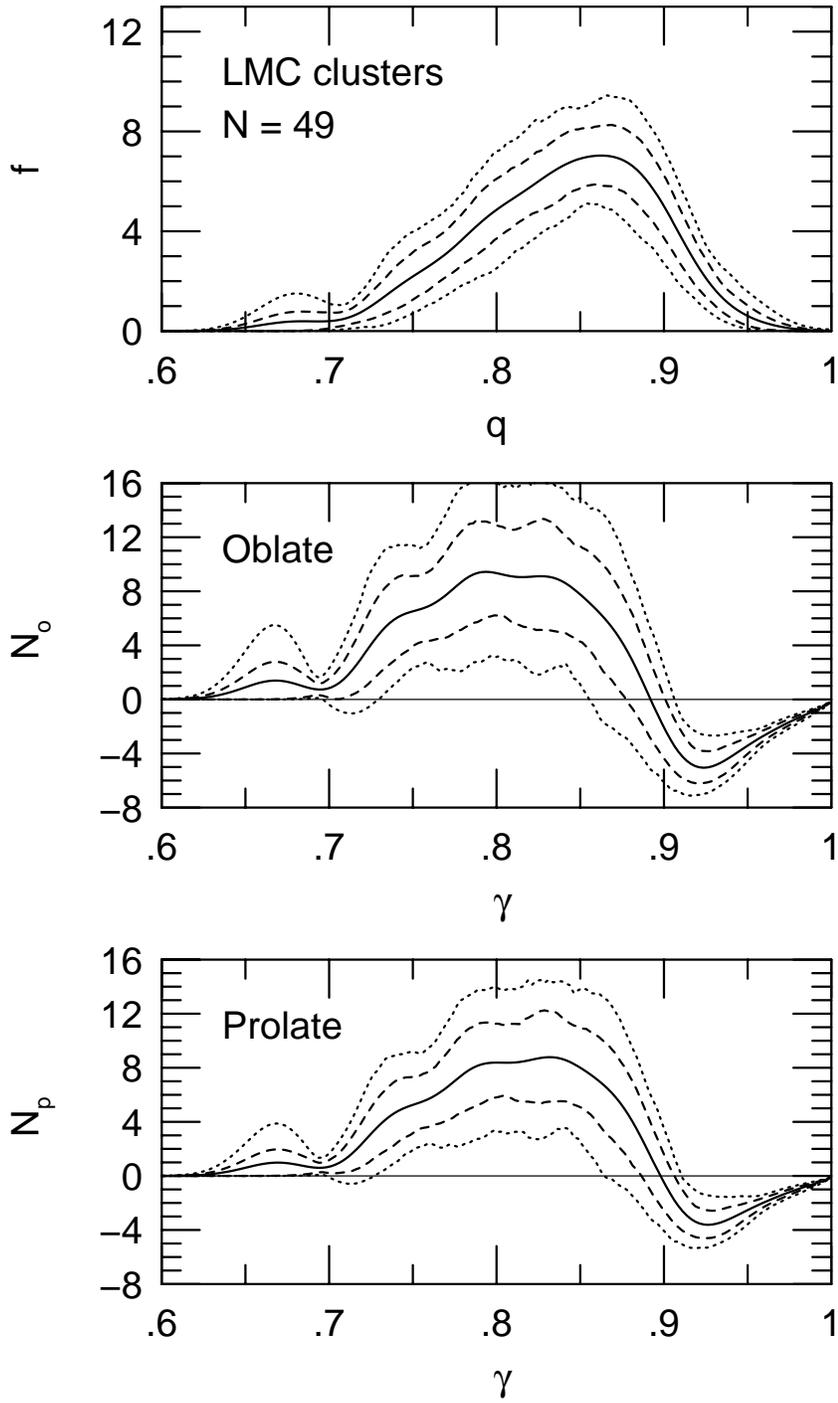

Fig. 2

– 20 –

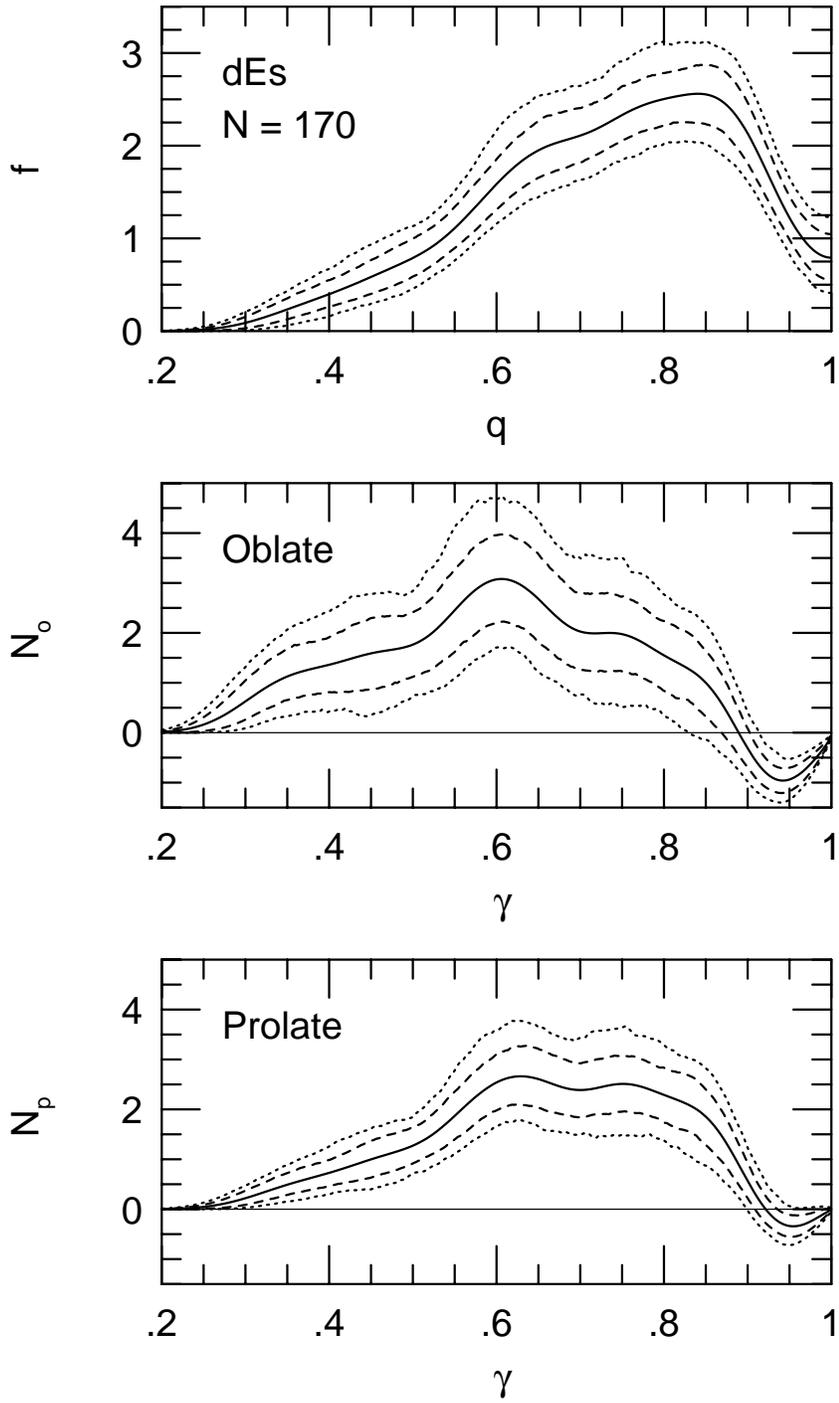

Fig. 3

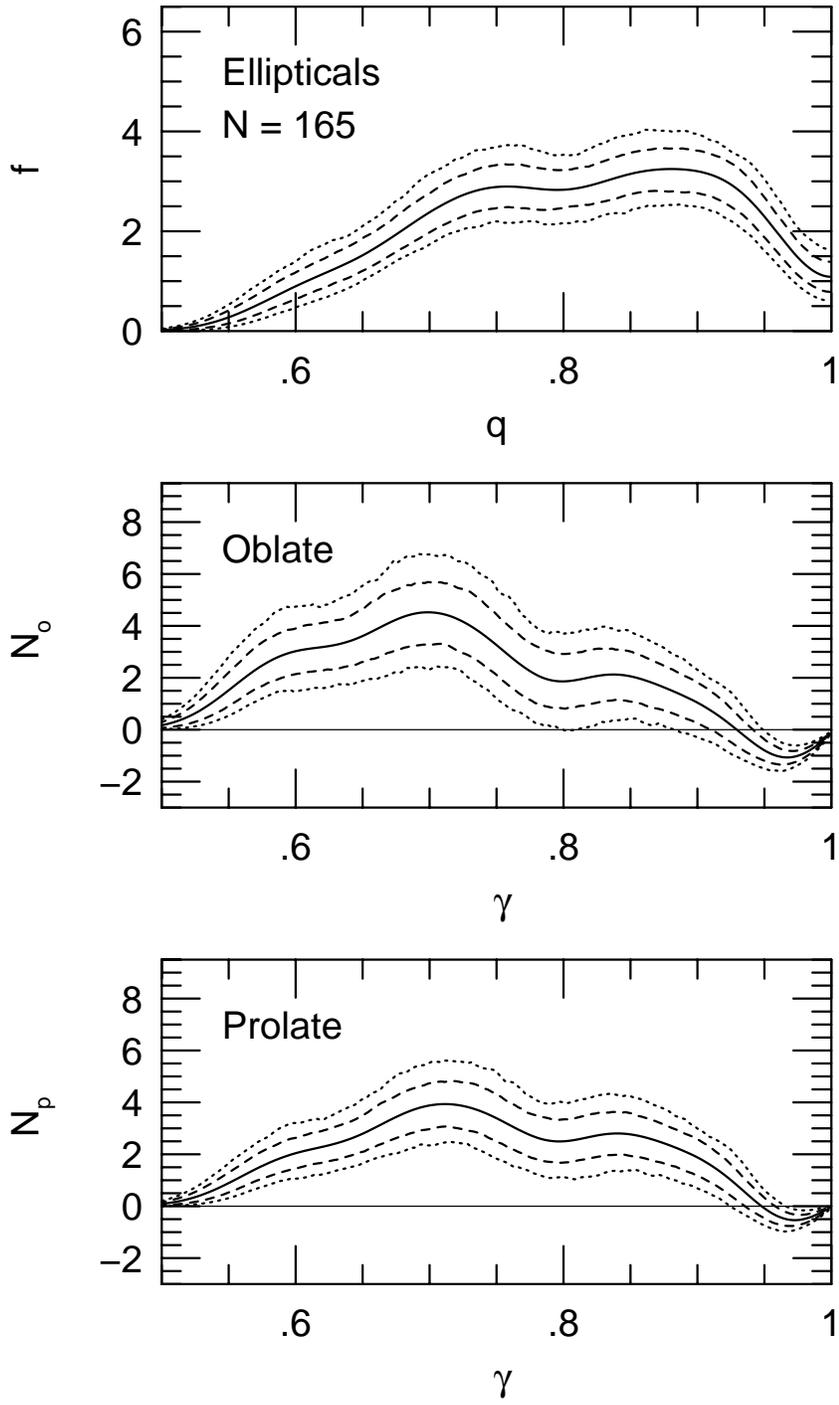

Fig. 4

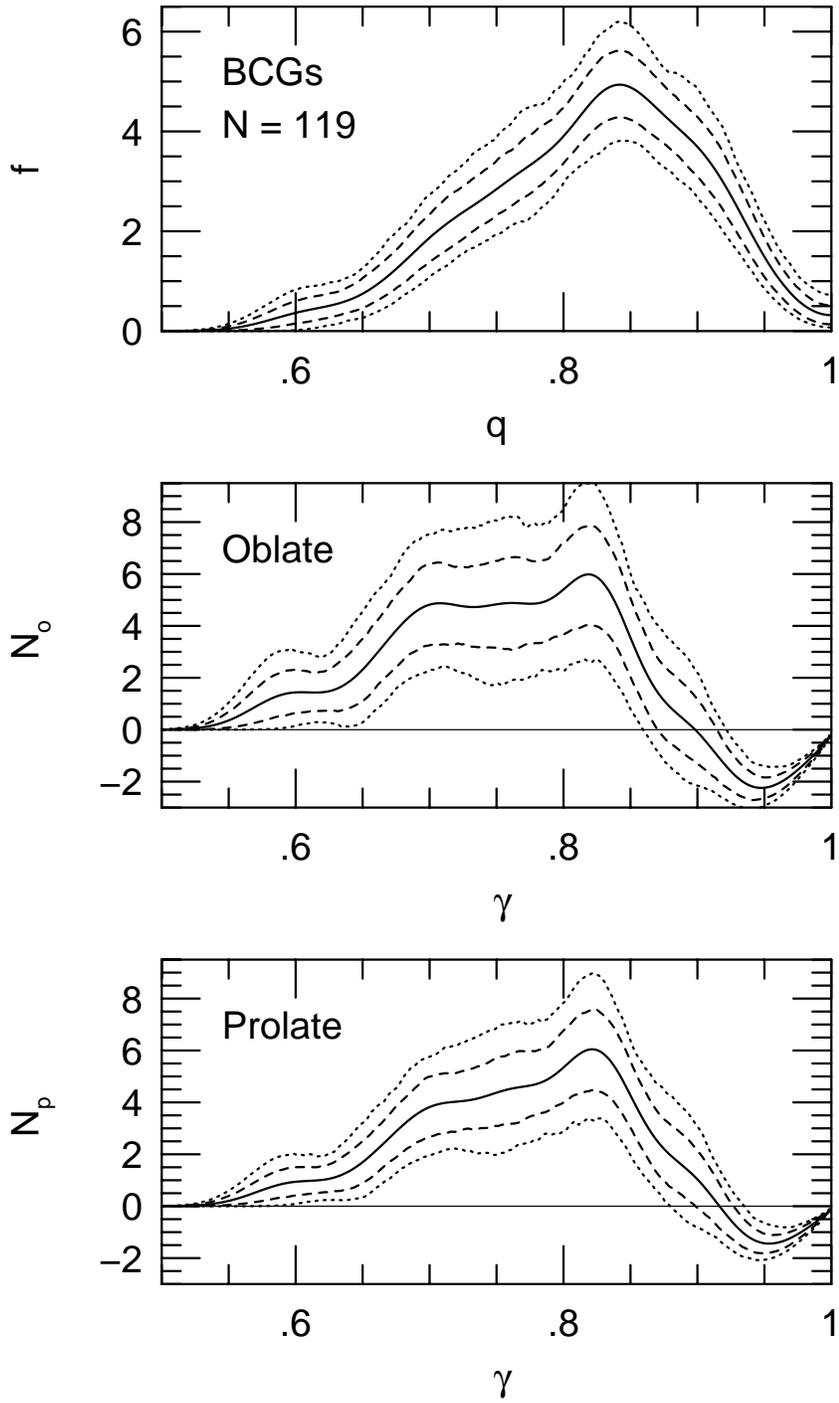

Fig. 5

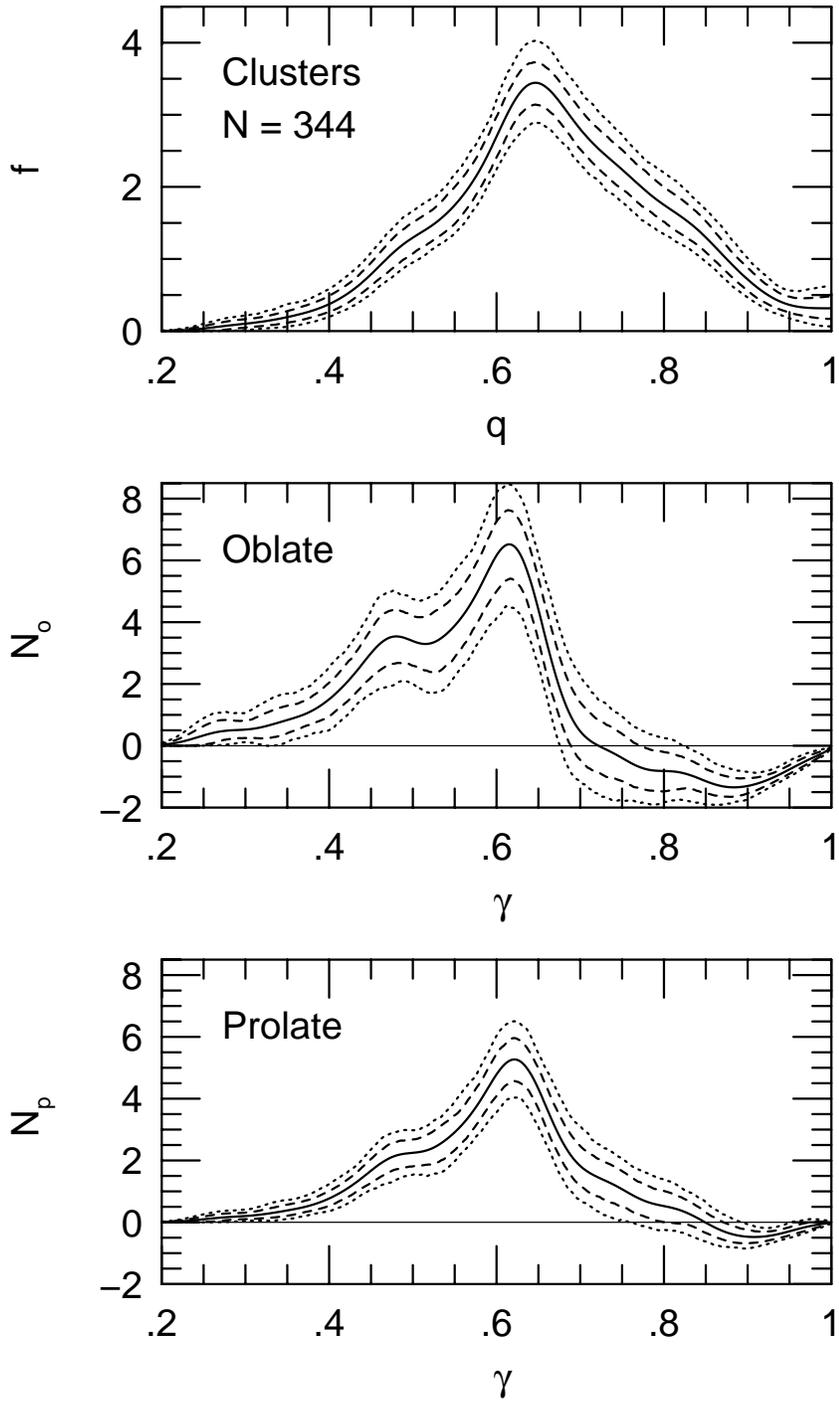

Fig. 6